\documentclass[runningheads]{llncs}
\usepackage{graphicx}
\usepackage{times}
\usepackage{latexsym}
\usepackage{todonotes}
\usepackage{times}
\usepackage{makecell}
\usepackage{url}
\usepackage{enumitem}
\usepackage{flushend}
\usepackage{latexsym}
\usepackage{makecell}
\usepackage{graphicx}
\usepackage{multirow}
\usepackage{subcaption,booktabs}
\usepackage{mathdots}
\usepackage{amsmath}
\usepackage{amsfonts}
\usepackage{amssymb}
\usepackage{bm}
\usepackage{calrsfs}
\usepackage[mathscr]{euscript}
\usepackage{array}

\usepackage[utf8]{inputenc}
\usepackage{mathtools, nccmath}

\usepackage{xpatch}
\xpatchcmd{\NCC@ignorepar}{%
\abovedisplayskip\abovedisplayshortskip}
{%
\abovedisplayskip\abovedisplayshortskip%
\belowdisplayskip\belowdisplayshortskip}
{}{}

\usepackage{blindtext}
\usepackage[group-separator={,},group-minimum-digits={3}]{siunitx}
\usepackage{scrextend}
\addtokomafont{labelinglabel}{\bfseries}%\sffamily
\usepackage{fancyvrb}
\newbox\verbbox

\RequirePackage{colortbl}

\definecolor{airforceblue}{rgb}{0.36, 0.54, 0.66}
\definecolor{amaranth}{rgb}{0.9, 0.17, 0.31}
\definecolor{applegreen}{rgb}{0.55, 0.71, 0.0}
\definecolor{alizarin}{rgb}{0.82, 0.1, 0.26}
\definecolor{azure}{rgb}{0.0, 0.5, 1.0}
\definecolor{cadmiumgreen}{rgb}{0.0, 0.42, 0.24}
\definecolor{dkgreen}{rgb}{0,0.6,0}
\definecolor{gray}{rgb}{0.5,0.5,0.5}
\definecolor{mauve}{rgb}{0.58,0,0.82}

\usepackage{microtype}
\usepackage[multiple]{footmisc}
\usepackage[hyperfootnotes=false]{hyperref}
\begin{document}
\title{ICDAR 2021 Competition on Scientific Table Image Recognition to LaTeX}

\author{Pratik Kayal\inst{1} \and
Mrinal Anand\inst{2} \and
Harsh Desai\inst{3} \and
Mayank Singh\inst{4}}
\authorrunning{Kayal \textit{et al.}}
\institute{Indian Institute of Technology, Gandhinagar, India \\ \email{pratik.kayal@iitgn.ac.in}
\and
Indian Institute of Technology, Gandhinagar, India \\ \email{mrinal.anand@iitgn.ac.in}\\
\and
Indian Institute of Technology, Gandhinagar, India  \\ \email{hsd31196@gmail.com}\\
\and
Indian Institute of Technology, Gandhinagar, India  \\ \email{singh.mayank@iitgn.ac.in}}

\maketitle 
\begin{abstract}
Tables present important information concisely in many scientific documents.
Visual features like mathematical symbols, equations, and spanning cells make structure and content extraction from tables embedded in research documents difficult.
This paper discusses the dataset, tasks, participants' methods, and results of the ICDAR 2021 Competition on Scientific Table Image Recognition to LaTeX. 
Specifically, the task of the competition is to convert a tabular image to its corresponding \LaTeX~source code.
% Further analysis demonstrates that the proposed models efficiently identify the number of rows and columns, the alphanumeric characters, the \LaTeX~tokens, and symbols.
We proposed two subtasks. In Subtask 1, we ask the participants to reconstruct the \LaTeX~structure code from an image. In Subtask 2, we ask the participants to reconstruct the \LaTeX~content code from an image.
% Three teams participated in TSR task and two teams participated in the TCR task. 
This report describes the datasets and ground truth specification, details the performance evaluation metrics used, presents the final results, and summarizes the participating methods.
Submission by team VCGroup got the highest Exact Match accuracy score of 74\% for Subtask 1 and 55\% for Subtask 2, beating previous baselines by 5\% and 12\%, respectively.
Although improvements can still be made to the recognition capabilities of models, this competition contributes to the development of fully automated table recognition systems by challenging practitioners to solve problems under specific constraints and sharing their approaches; the platform will remain available for post-challenge submissions at \href{https://competitions.codalab.org/competitions/26979}{https://competitions.codalab.org/competitions/26979}.
\end{abstract}
\section{Introduction}
\label{introduction}
% Tables are ubiquitous in written reports from private, public, and governmental institutions. 
Scientific documents contain tables that hold meaningful information such as final results of an experiment or comparisons with earlier baselines in a concise manner, but accurately extracting the information proves to be a challenging task~\cite{niklaus2018survey,singh2019automated}.
In recent years, we witness the widespread usage of several typesetting tools for document generation. In particular, tools like \LaTeX~ help in typesetting complex scientific document styles and subsequent electronic exchange documents.
It has been shown in a study~\cite{brischoux2009don} that 26\% of submissions to 54 randomly selected scholarly journals from 15 different scientific disciplines use \LaTeX~typesetter, and there is a significant difference between \LaTeX-using and non-\LaTeX-using disciplines.
Tables can be presented in a variety of ways visually, such as spanning multiple columns or rows, with or without horizontal and vertical lines, non-standard spacing and alignment, and text formatting~\cite{embley2006table}.
The table detection task involves predicting the bounding boxes around tables embedded within a PDF document, while the table recognition task involves the table structure and content extraction. We find significantly large number of datasets for table detection and table structure recognition~\cite{fang2012dataset,gao2019icdar,li2020tablebank,siegel2018extracting,zhong2019publaynet}. Comparatively, table content recognition is a least explored task with a very few datasets~\cite{8978078,gobel2013icdar,zhong2019imagebased}. 
Interestingly, the majority of these datasets are insufficient to perform end-to-end neural training. For instance, ICDAR 2013 table competition task~\cite{gobel2013icdar} includes only 150 table instances. Besides, several datasets comprise simpler tables that might not generalize in real-world extraction scenarios. 
Tables written in \LaTeX~utilizes libraries like \textit{booktabs, array, multirow, longtable}, and \textit{graphicx}, for table creation and libraries like \textit{amsmath, amssymb, amsbsy}, and \textit{amsthm}, for various scientific and mathematical writing. Such libraries help in creating complex scientific tables that are difficult to recognize by the current systems accurately~\cite{singh2019automated}.  Deng et al.~\cite{8978078} proposed a table recognition method for \LaTeX~generated documents. However, due to minimal data processing like word-level tokenization and no fine-grained task definitions, they predict the entire \LaTeX~code leading to poor recognition capabilities.

\begin{figure}[tb]
	\begin{subtable}[h]{0.487\textwidth}
\resizebox{\hsize}{!}{
\begin{tabular}{lc}

\rotatebox[origin=c]{90}{\centering\textsc{\textbf{Table}}} &

        \begin{tabular}{ | l || c | r | r | }
          \hline
           & $\epsilon_{LJ} \tiny{[\frac{Kcal}{mol} ]}$ & $\sigma_{LJ}$ \tiny{[\AA]} & $r_{cut}$ \tiny{[\AA]}\\
          \hline
         $C-C$ & 0.0951 & 3.473 & 15.0  \\
         $C-H$ & 0.0380 & 3.159 & 15.0  \\
         $H-H$ & 0.0152 & 2.846 & 15.0\\
         \hline
        \end{tabular}\\\addlinespace[0.4cm]
\end{tabular}}

% \rotatebox[origin=c]{90}{\centering\textsc{\textbf{OCR}}} &
% 		\begin{tabular}{ c }
%     		ensla—] | obs Al | Teut 1A \textbackslash\textbackslash n\\
    		
%             C-C 0.0951 3.473 15.0 \textbackslash\textbackslash n
%             \\
%             C'— HAH 0.0380 3.159 15.0\textbackslash\textbackslash n
%             \\
%             H—-H 0.0152 2.846 15.0
%         \end{tabular}
% \end{tabular}}
		\caption{Scientific table in ~\cite{tsourtis2017parameterization}.}
	\end{subtable}
	\hfill
	\begin{subtable}[h]{0.5\textwidth}
\resizebox{\hsize}{!}{
\begin{tabular}{lc}

\rotatebox[origin=c]{90}{\centering\textsc{\textbf{Table}}} &
    	\begin{tabular}{c|ccc}
                & \multicolumn{3}{c}{Search Strategies}\\
                & (RMSE, $\sigma$) & & (PAcc, $\sigma$)\\ \hline\hline
        SkILL & (0.616, 0.063) &  & (0.661, 0.045)  \\
        SkILL+pruning & (0.581, 0.099)  & & \textbf{(0.663, 0.045)}  \\\hline
        Aleph & \multicolumn{3}{c}{(0.656, 0.047)}\\
        \end{tabular}\\\addlinespace[0.3cm]

% \rotatebox[origin=c]{90}{\centering\textsc{\textbf{OCR}}} &
% 		\begin{tabular}{c}
%     		Search Strategies\textbackslash\textbackslash n
%             (RMSE, oc) \\(PAcc, co)\textbackslash\textbackslash n\textbackslash\textbackslash n\textbackslash\textbackslash n
%             SkILL\textbackslash\textbackslash n
%             SkILL+pruning\\\textbackslash\textbackslash n\textbackslash\textbackslash n
%             (0.616, 0.063) (0.661, 0.045)\textbackslash\textbackslash n\\\textbackslash\textbackslash n
%             (0.581, 0.099) (0.663, 0.045)\\\textbackslash\textbackslash n
%             Aleph\textbackslash\textbackslash n\textbackslash\textbackslash n\textbackslash\textbackslash n
%             (0.656, 0.047)\\
%         \end{tabular}
\end{tabular}}
		\caption{Scientific table in ~\cite{corte2015skill}.}
	\end{subtable}
	\caption{Examples of naturally occurring scientific tables.}
	\label{tesseract}
\end{figure}

\noindent To understand and tackle this challenging problem of table recognition, we organized a competition with the objective to convert a given image of the table to its corresponding (i) structure \LaTeX~code and (ii) content \LaTeX~code. 
This report presents the final results after analyzing the submissions received.

\noindent\textbf{The paper outline :} The entire paper is organized as follows. Section~\ref{competition_organization} describes the organization of the competition. Section~\ref{tasks} describes the tasks descriptions. Section~\ref{dataset} describes the dataset used in the competition. Section~\ref{evaluation} details about the performance evaluation metrics. Section~\ref{methods} presents the description of baselines and methods submitted by the participants. Section~\ref{sec:results} presents the final results and discussions.

\section{Competition Organization}
\label{competition_organization}
The Competition on Scientific Table Image Recognition to \LaTeX~was organized with the objective to evaluate structure recognition capabilities and content recognition capabilities of the current state-of-the-art systems. 
We use the Codalab\footnote{\url{https://competitions.codalab.org/}} platform for submission and automatic evaluation of methods. Multiple submissions were allowed for each team. After each submission, evaluation metrics were computed, and the leaderboard was updated accordingly based on the highest score. 
The competition ran between October 2020 and March 2021 and was divided into three phases, validation, testing, and post-evaluation. In the validation phase, participants were given training and validation dataset and were allowed to test their system on the validation set. In the testing phase held in March 2021, participants submitted their solutions on the test dataset on which final results are evaluated. For the testing phase, the leaderboard was hidden from the participants. In the post-evaluation phase, participants are allowed to submit their solutions and improve their rankings on the leaderboard. 

% The competition consists of two tasks: Table structure reconstruction (Task I), Table content reconstruction (Task II). The objective is to convert a given image of the table to its corresponding (i) structure \LaTeX~code for Task I and (ii) content \LaTeX~code for task II. The participants were allowed multiple submissions per day for both tasks.

\section{Task Description} 
\label{tasks}
The task of generating \LaTeX~code from table images is non-trivial. Similar features of the table can be encoded in different ways in the \LaTeX~code. While the content in a tabular image is in sequential form, structural information is encoded in the form of structural tokens like \verb+\multicolumn+ and column alignment tokens like \verb+c+, \verb+l+, and \verb+r+ in \LaTeX. The proposed recognition models generate a sequence of tokens given the input table image. We examine the task of generating \LaTeX~code from table image by dividing it into two sub-tasks --- \textbf{Table Structure Recognition} and \textbf{\LaTeX~Optical Character Recognition}. The extracted structure information aids in reconstructing the table structure. In comparison, the extracted content information can help fill the textual placeholders of the table.

\subsection{Task I: Table Structure Reconstruction(TSR)}
\begin{figure*}[!t]
\centering \includegraphics[width=\linewidth]{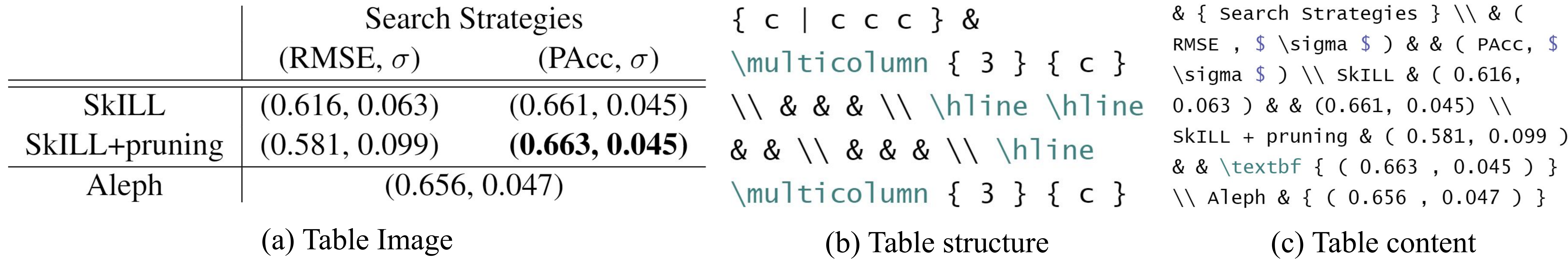}
\caption{Sample of the dataset. The words in table content (except \LaTeX~tokens) are recognized as individual characters and then connected using a separate post-processing method. (a) Input to the models, (b) Sample with structure tokens, (c) Sample with content tokens.}
\label{fig:data_prep}
\vspace{-0.15in}
\end{figure*}

This subtask recognizes the structural information embedded inside the table images. Figure~\ref{fig:data_prep}b shows a TSR output from a sample  tabular image (see Figure~\ref{fig:data_prep}a).
Structural information such as mutliple columns (defined using the command \verb+\multicolumn}+ \verb+{cols}{pos}{text}+), multiple rows (defined using the command \verb+\multirow+ \verb+{rows}{width}{text}+),  the column alignment specifiers (\verb+c+, \verb+l+, or \verb+r+), horizontal lines (\verb+\hline+, \verb+\toprule+, \verb+\midrule+, or \verb+\bottomrule+), etc., are recognized in this task. Note that, content inside the third argument (\verb+{text}+) of \verb+\multicolumn+ or \verb+\multirow+  commands are recognised in L-OCR task (defined in the next section).
A special placeholder token \textit{`CELL'} represents a content inside a specific cell of the table. Specifically, the vocabulary of TSR task comprises \verb+&+, \verb+0+, \verb+1+, \verb+2+, \verb+3+, \verb+4+, \verb+5+, \verb+6+, \verb+7+, \verb+8+, \verb+9+, \verb+CELL+, \verb+\\+, \verb+\hline+, \verb+\hspace+, \verb+\multirow+, \verb+\multicolumn+, \verb+\toprule+, \verb+\midrule+, \verb+\bottomrule+, \verb+c+, \verb+l+, \verb+r+, \verb+|+, \verb+\{+, and \verb+\}+. 
% Figure~\ref{fig:data_prep} shows the table image and its corresponding \LaTeX~code representing structure information.

% \begin{figure}[h]
% \centering
%     \subfloat[\centering Table Image]{{\includegraphics[width=0.5\textwidth]{main/figures/table_image_example.jpg} }}%
%     \qquad
%     \subfloat[\centering Table Structure Information ]{{\includegraphics[width=0.5\textwidth]{main/figures/table_structure.png} }}%
%     % \qquad
%     % \subfloat[\centering Content Information ]{{\includegraphics[width=0.5\textwidth]{figures/content.png} }}%
%     \caption{Example of table image along with its structure information.}
%     \label{fig:table_example1}%
% \end{figure}

\subsection{Task II: Table Content Reconstruction(TCR)}
This subtask extracts content information from the table images. Content information includes alphanumeric characters, symbols, and \LaTeX~tokens. While the table's basic structure is recognized in the TSR task, the more nuanced structural information is captured by the L-OCR task. Instead of predicting all possible tokens leading to vocabulary size in millions, we demonstrate a character-based recognition scheme. A delimiter token (\verb+¦+) is used to identify words among the characters. Figure~\ref{fig:data_prep}c shows the TCR output from a sample tabular image (see Figure~\ref{fig:data_prep}a). The terms \verb+(Accuracy, (%))+ together produces the first cell in the first row of the table. \verb+Accuracy+ is recognized as \verb+A+ \verb+<dim>+ \verb+c+ \verb+<dim>+ \verb+c+ \verb+<dim>+ \verb+u+ \verb+<dim>+ \verb+r+ \verb+<dim>+ \verb+a+ \verb+<dim>+ \verb+c+ \verb+<dim>+ \verb+y+. The keyword \verb+<dim>+ (delimiter token) is removed in the post-processing step. Overall, 235 unique tokens comprises the vocabulary of L-OCR sub-task. This includes all the alphabets and their case variants (a-z and A-Z), digits (0-9), \LaTeX~environment tokens (\verb+\vspace+, \verb+\hspace+, etc.), brackets (\verb+{+, \verb+}+, etc.), modifier character (\verb+\+), accents (\verb+,+, \verb+^+, etc.) and symbols (\verb+%+, \verb+#+, \verb+&+, etc.).

% \begin{figure}[h]
% \centering
%     \subfloat[\centering Table Image]{{\includegraphics[width=0.5\textwidth]{main/figures/table_image_example.jpg} }}%
%     \qquad
%     \subfloat[\centering Table Content Information ]{{\includegraphics[width=0.5\textwidth]{main/figures/table_content.png} }}%
%     % \qquad
%     % \subfloat[\centering Content Information ]{{\includegraphics[width=0.5\textwidth]{figures/content.png} }}%
%     \caption{Example of table image along with its content information.}
%     \label{fig:table_example2}%
% \end{figure}

\section{Dataset}
\label{dataset}
% Due to the unavailability of the publicly available corpus for the generation of \LaTeX~code from table images, we create a new dataset for Table image to Latex. 
The dataset contains table images and its corresponding \LaTeX~code representing structure and content information. 
We only consider papers corresponding to the topics in Computer Science, from the preprint repository \textit{ArXiv}\footnote{\url{http://arxiv.org/}}.  We preprocess \LaTeX~code before compiling them to images and post-process the \LaTeX~code to make them more suitable for the text generation task.
The \LaTeX~source code is processed to generate table images and post-processed to get corresponding ground truth representation. We extract the table snippet during preprocessing, which begins with \verb+\begin{tabular}+ command and ends with \verb+\end{tabular}+ command. We remove commands like \verb+\cite{}+, \verb+\ref{}+ along with the commented text that cannot be predicted from the tabular images. The filtered tabular \LaTeX~code is compiled as PDF and then converted into JPG images with 300dpi resolution using Wand library\footnote{\url{https://github.com/emcconville/wand}}. To keep a smaller vocabulary for faster training, we mask \LaTeX~environment (beginning with \verb+\+ character) tokens with corpus frequency less than 1000 as a special token \verb+\LATEX_TOKEN+.
We separate the structure and content-related token of the table from the \LaTeX~source code during post-processing to create the dataset for TSR and TCR tasks, respectively. The structural tokens contain tabular environment parameters, spanning rows and columns tokens, horizontal and line tokens. The content tokens include alphanumeric characters, mathematical symbols, and other \LaTeX~environment tokens. Dataset for the TSR task and TCR task is created by considering ground truth \LaTeX~code with a maximum length of 250 and 500, respectively.

\begin{table}[t]
	\centering
% 	\resizebox{\hsize}{!}{
    % \footnotesize
    % \setlength{\tabcolsep}{12pt}
	\begin{tabular}{|l|c|c|c|c|c|c|}
		\hline
        \multicolumn{1}{|l|}{\textbf{Dataset}} &
		\multicolumn{1}{|c}{\textbf{ML}} &
		\multicolumn{1}{|c}{\textbf{Samples}} &
        \multicolumn{1}{|c}{\textbf{Train}} &
        \multicolumn{1}{|c}{\textbf{Val}} &
        \multicolumn{1}{|c}{\textbf{Test}} &
        \multicolumn{1}{|c|}{\textbf{T/S}}   \\ \hline
        TSRD & 250 & \num{46141}  & \num{43138} & \num{800} & \num{2203} & 76.09 \\ \hline
  		TCRD & 500 & \num{37917} & \num{35500} & \num{500} & \num{1917} & \num{213.95} \\ \hline
	\end{tabular}
% 	}
  	\caption{Summary of datasets. ML denotes the maximum sequence length of target sequence. Samples denote the total number of image-text pairs. T/S is the average number of tokens per sample.}
  	\label{tab_datasets}
\end{table}

\section{Evaluation metrics}
\label{evaluation}
We use exact match accuracy and exact match accuracy @ 95\% similarity as our core evaluation metrics for evaluating both tasks.
Consider that if the model generates a hypothesis $\bm{y^{\ast}}$ consisting of tokens $y^{\ast}_1 y^{\ast}_2 \cdots y^{\ast}_{n_1}$, during the test phase. We evaluate by comparing it with the ground truth token sequence code $\bm{y}$ consisting of tokens $y_1 y_2 \cdots y_{n_2}$, where $n_1$ and $n_2$ are the number of tokens in model-generated code and ground truth code, respectively. 
\begin{enumerate}
    \item \noindent \textbf{Exact Match Accuracy (EM):} For a correctly identified sample, $y^{\ast}_1=y_1, y^{\ast}_2=y_2, \cdots, y^{\ast}_{n_1}=y_{n_2}$.
    \item \noindent \textbf{Exact Match Accuracy @95\% (EM @95\%):} For a correctly identified sample, $y^{\ast}_i=y_j, y^{\ast}_{i+1}=y_{j+1}, \cdots, y^{\ast}_{i+l}=y_{j+l}$ and $l>=0.95\times n_2$. In this metric there are no insertions/deletions allowed in between and move is towards (i+1, j+1) from (i,j).
\end{enumerate}

In addition, we also employ task-specific metrics that provide intuitive criteria to compare different models trained for the specific task. For the TSR task, we use row accuracy and column accuracy as evaluation metrics.
\begin{enumerate}
    \item \noindent \textbf{Row Accuracy (RA):} For a correctly identified sample, the number of rows in $\bm{y^{\ast}}$ is equal to the number of rows in $\bm{y}$.
    \item \noindent \textbf{Column Accuracy (CA):} For a correctly identified sample, the number of columns in $\bm{y^{\ast}}$ is equal to the number of columns in $\bm{y}$.
\end{enumerate}

We use Alpha-Numeric characters prediction accuracy, \LaTeX~Token accuracy, \LaTeX~symbol accuracy, and  Non-\LaTeX~ symbols prediction accuracy for the TCR task.
Each token in $\bm{y^{\ast}}$ and $\bm{y}$ is classified into four exhaustive categories: (i) Alpha-Numeric Tokens (Alphabets and Numbers), (ii) \LaTeX~Tokens (tokens that are defined in and used by \LaTeX~markup language like \verb+\cdots+, \verb+\times+, \verb+\textbf+, etc.), (iii) \LaTeX~Symbols (symbols with escape character (`\textbackslash') like \verb+\%+, \verb+\$+, \verb+\{+, \verb+\}+, etc.), and (iv) Non-\LaTeX~Symbols (symbols like \verb+=+, \verb+$+, \verb+{+, \verb+}+, etc.). 

\begin{enumerate}
    \item \noindent \textbf{Alpha-Numeric Tokens Accuracy (AN):} We form strings $\bm{y^{\ast}_{AN}}$ and $\bm{y_{AN}}$ from  $\bm{y^{\ast}}$ and $\bm{y}$, respectively, by keeping the alpha-numeric tokens and discarding the rest and preserving the order. Then, for a correctly identified sample, $\bm{y^{\ast}_{AN}} = \bm{y_{AN}}$.
    \item \noindent \textbf{\LaTeX~Tokens Accuracy (LT):} We form strings $\bm{y^{\ast}_{LT}}$ and $\bm{y_{LT}}$ from  $\bm{y^{\ast}}$ and $\bm{y}$, respectively, by keeping the \LaTeX~tokens and discarding the rest and preserving the order. Then, for a correctly identified sample, $\bm{y^{\ast}_{LT}} = \bm{y_{LT}}$.
    \item \noindent \textbf{\LaTeX~Symbols  Accuracy (LS):} We form strings $\bm{y^{\ast}_{LS}}$ and $\bm{y_{LS}}$ from  $\bm{y^{\ast}}$ and $\bm{y}$, respectively, by keeping the \LaTeX~symbol tokens and discarding the rest and preserving the order. Then, for a correctly identified sample, $\bm{y^{\ast}_{LS}} = \bm{y_{LS}}$.
    \item \noindent \textbf{Non-\LaTeX~Symbols Accuracy (NLS):} We form strings $\bm{y^{\ast}_{NLS}}$ and $\bm{y_{NLS}}$ from  $\bm{y^{\ast}}$ and $\bm{y}$, respectively, by keeping the non-\LaTeX~symbol tokens and discarding the rest and preserving the order. Then, for a correctly identified sample, $\bm{y^{\ast}_{NLS}} = \bm{y_{NLS}}$.
\end{enumerate}

\section{Baseline and Participating Methods}
\label{methods}
\subsection{CNN-Baseline} The image-to-markup model proposed by Deng et al.~\cite{deng2017image} extracts image features using a convolutional neural network (CNN) and arranges the features in a grid. CNN consists of eight convolutional layers with five interleaved max-pooling layers. Each row is then encoded using a bidirectional LSTM with the initial hidden state (called positional embeddings) kept as trainable to capture semantic information in the vertical direction. An LSTM decoder then uses these encoded features with a visual attention mechanism to predict the tokens given the previously generated token history and the features as input.

\subsection{Transformer-Baseline} Recently, Transformer~\cite{vaswani2017attention} based architectures have shown state-of-the-art performance for several Image-to-Text tasks~\cite{lyu20192d,yang2019simple}. We utilize a ResNet-101~\cite{he2015deep} layer that encodes the table image and a Transformer model for text generation. This model is based on architecture proposed in~\cite{feng2020scene}.

\subsection{VCGroup} Their model is based on MASTER~\cite{DBLP:journals/corr/abs-1910-02562} originally developed for scene recognition task. MASTER uses a novel multi-aspect non-local block and fuses it into the conventional CNN backbone, enabling the feature extractor to model a global context. The proposed multi-aspect non-local block can learn different aspects of spatial 2D attention, which can be viewed as a multi-head self-attention module. The proposed solution is a highly optimized version of the original MASTER model. The task wise optimizations and techniques used by them are explained below:

\begin{table}[h]
	\centering
% 	\resizebox{\hsize}{!}{
    % \footnotesize
    % \setlength{\tabcolsep}{12pt}
	\begin{tabular}{|l|c|c|c|c|c|}
		\hline
        \multicolumn{1}{|l|}{\textbf{Method}} &
		\multicolumn{1}{c}{\textbf{EM}} &
		\multicolumn{1}{|c}{\textbf{EM @95\%}} &
        \multicolumn{1}{|c}{\textbf{RA}} &
        \multicolumn{1}{|c|}{\textbf{CA}}   \\ \hline
        VCGroup & 0.74 & \num{0.88}  & \num{0.95} & \num{0.89}  \\ \hline
        Transformer-Baseline & 0.69 & 0.85 & 0.93 & 0.86 \\ \hline
        CNN-Baseline & 0.66 & 0.79 & 0.92 & 0.86 \\ \hline
        Format* & 0.57 & \num{0.80} & \num{0.91} & \num{0.87} \\ \hline
        asda* & 0.50 & \num{0.75} & \num{0.90} & \num{0.86} \\ \hline
	\end{tabular}
	\caption{Table Structure Reconstruction Results. *Method description unavailable.}
  	\label{result_I}
% 	}
  	
\end{table}

\begin{table}[h]
	\centering
% 	\resizebox{\hsize}{!}{
    % \footnotesize
    % \setlength{\tabcolsep}{12pt}
	\begin{tabular}{|l|c|c|c|c|c|c|}
		\hline
        \multicolumn{1}{|l|}{\textbf{Method}} &
		\multicolumn{1}{c}{\textbf{EM}} &
		\multicolumn{1}{|c}{\textbf{EM @95\%}} &
        \multicolumn{1}{|c}{\textbf{AN}} &
        \multicolumn{1}{|c|}{\textbf{LT}} &
        \multicolumn{1}{|c}{\textbf{LS}} &
        \multicolumn{1}{|c|}{\textbf{NLS}} \\ \hline
        VCGroup & 0.55 & \num{0.74}  & \num{0.85} & \num{0.75} & \num{0.96} & \num{0.62} \\ \hline
        Transformer-Baseline & 0.43 & 0.64 & 0.74 & 0.67 & 0.94 & 0.54 \\ \hline
        CNN-Baseline & 0.41 & 0.67 & 0.76 & 0.67 & 0.94 & 0.48 \\ \hline
        Format* & 0.0 & \num{0.52}  & \num{0.67} & \num{0.54} & \num{0.92} & \num{0.35} \\ \hline
	\end{tabular}
% 	}
	\caption{Table Content Reconstruction Results. *Method description unavailable.}
  	\label{result_II}
\end{table}

\begin{figure}[h]
\centering
    \subfloat[\centering TSR result ]{{\includegraphics[scale=0.4]{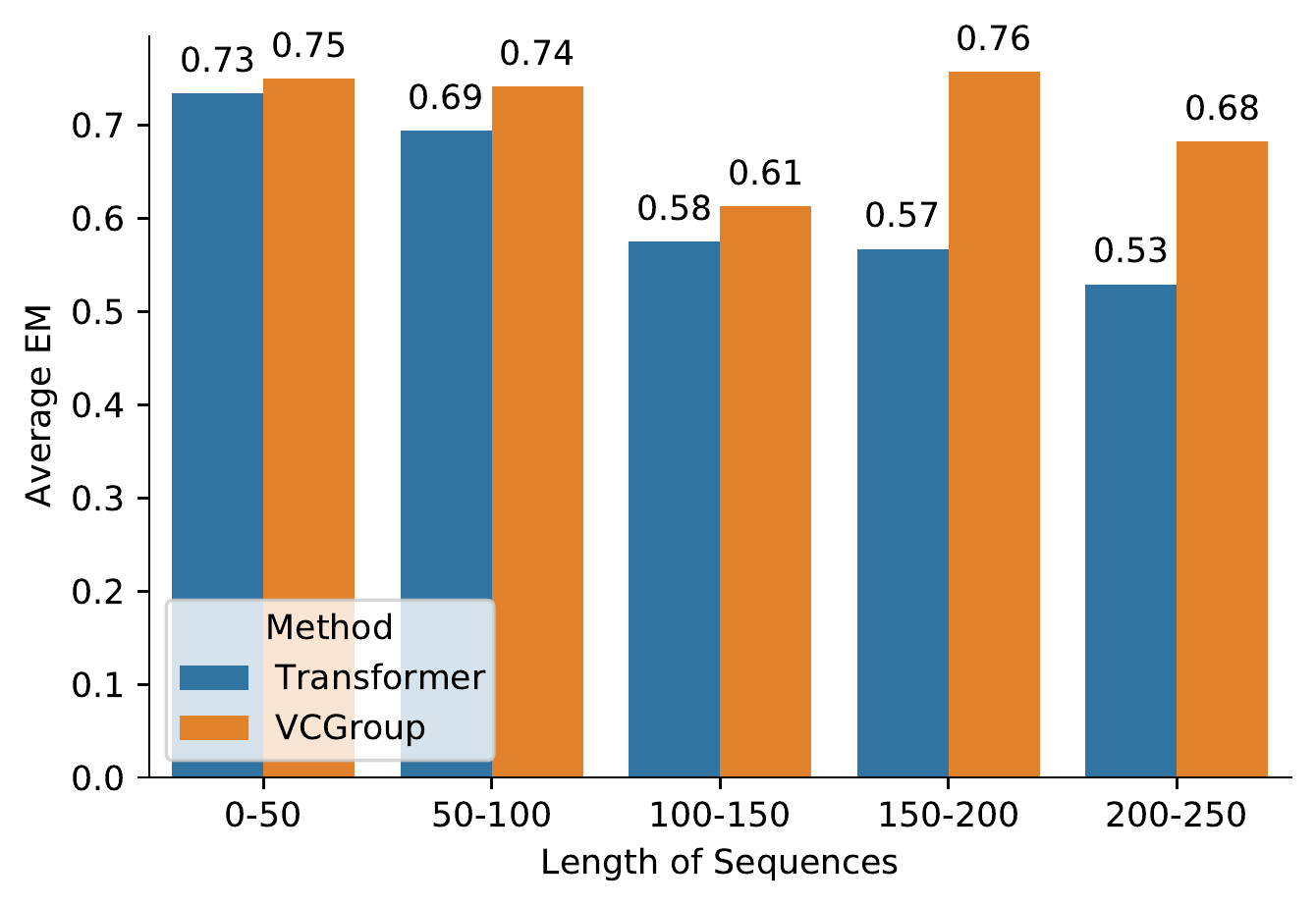} }}%
    \qquad
    \subfloat[\centering TCR result ]{{\includegraphics[scale=0.4]{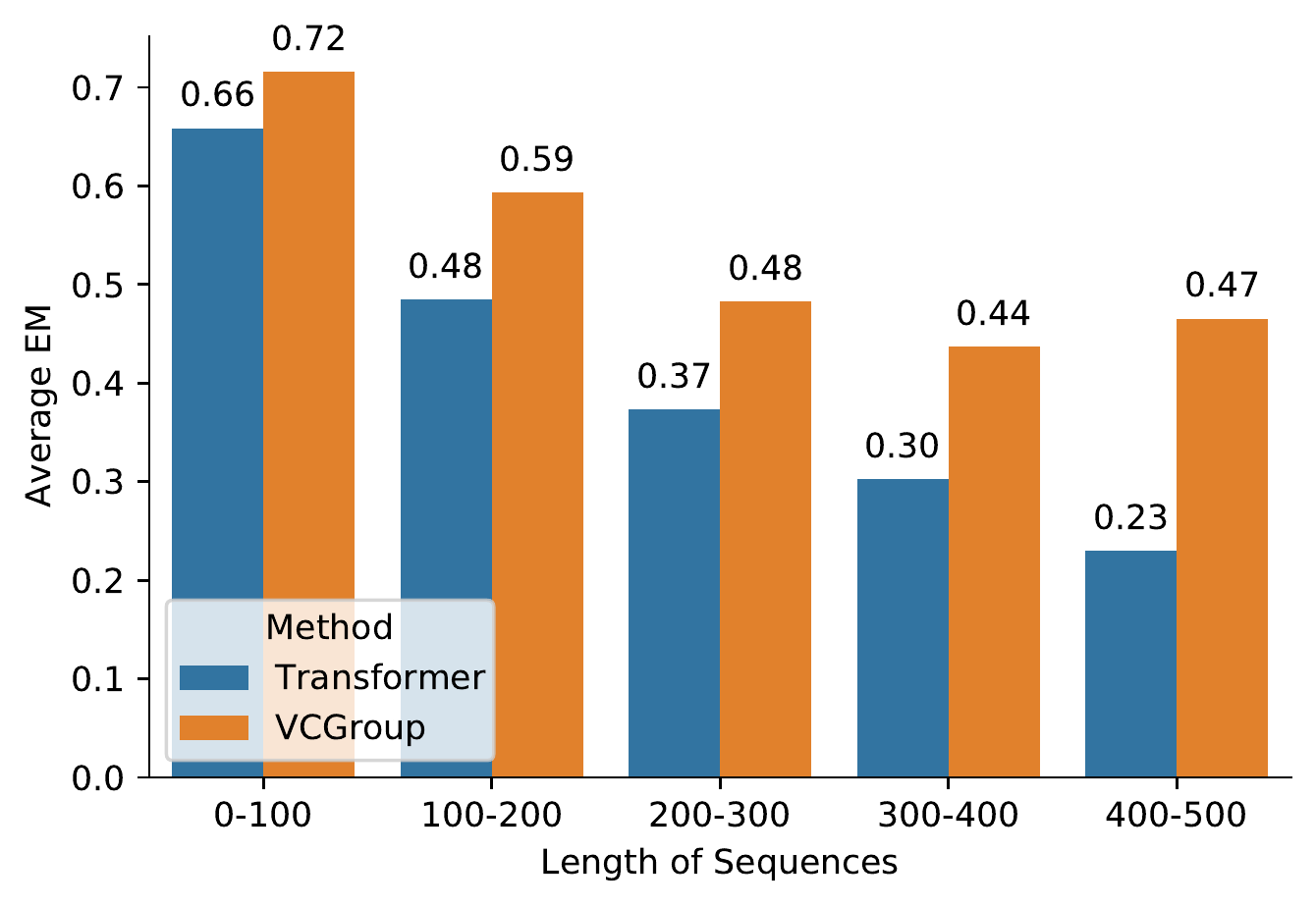} }}%
    \caption{Average EM score of VCGroup and Transformer-Baseline for both (a) TSR task and (b) TCR task.}
    \label{fig:avg_em}
\end{figure}

\begin{figure}[!h]
\centering
     \subfloat[\centering Example 1: Table Image ]{{\includegraphics[scale=0.25]{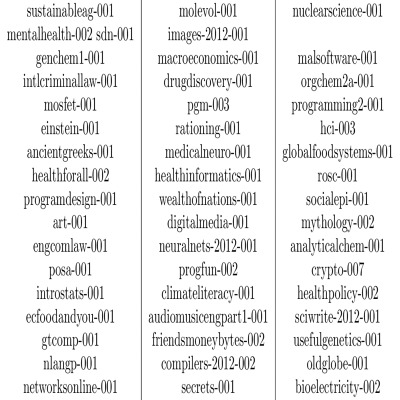} }}%
    \qquad
    \subfloat[\centering Example 1: Outputs ]{{\includegraphics[scale=0.5]{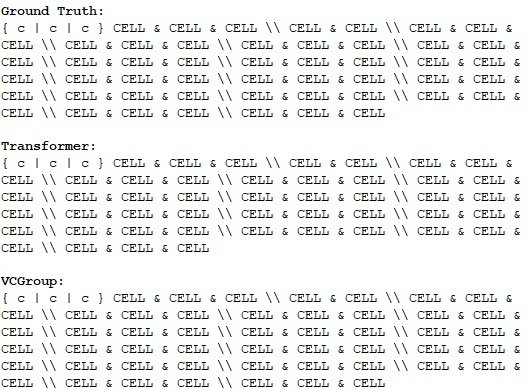} }}%
    \qquad
    \\
    \subfloat[\centering Example 2: Table Image ]{{\includegraphics[scale=0.25]{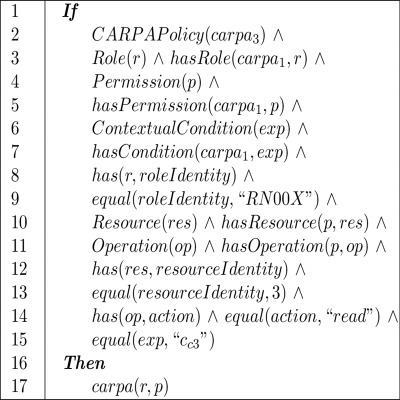} }}%
    \qquad
    \subfloat[\centering Example 2: Outputs ]{{\includegraphics[scale=0.5]{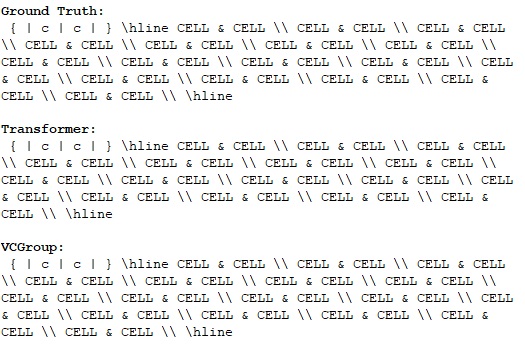} }}%
    \caption{Examples of correct cases by VCGroup method in the TSR task.}
    \label{fig:tsr_correct_results}%
\end{figure}

\begin{figure}[h]
\centering
    \subfloat[\centering Example 1: Table Image ]{{\includegraphics[scale=0.25]{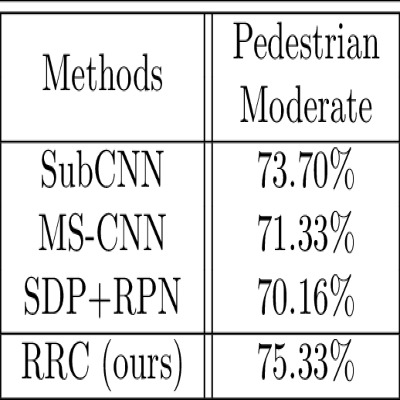} }}%
    \qquad
    \subfloat[\centering Example 1: Outputs ]{{\includegraphics[scale=0.5]{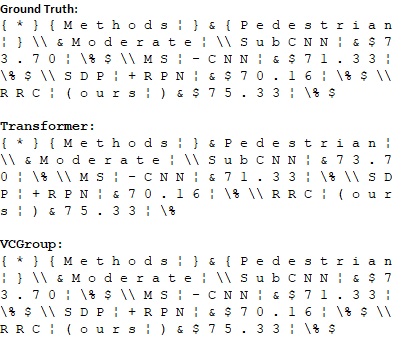} }}%
    \qquad
     \subfloat[\centering Example 2: Table Image ]{{\includegraphics[scale=0.25]{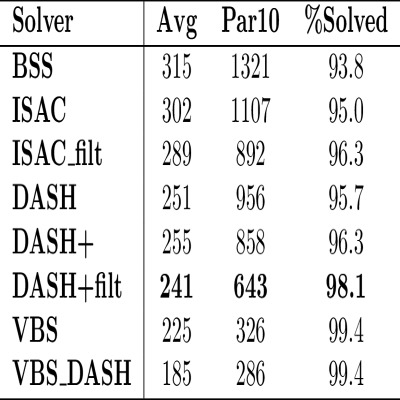} }}%
    \qquad
    \subfloat[\centering Example 2: Outputs ]{{\includegraphics[scale=0.5]{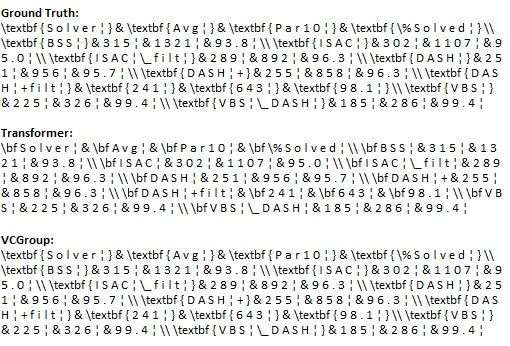} }}%
    
    \caption{Examples of correct cases by VCGroup method in the TCR task.}
    \label{fig:tcr_correct_results}%
\end{figure}

\begin{figure}[h]
\centering
    \subfloat[\centering Table Image ]{{\includegraphics[scale=0.25]{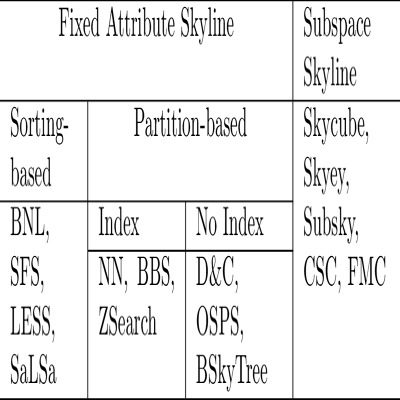} }}%
    \qquad
    \subfloat[\centering Outputs ]{{\includegraphics[scale=0.5]{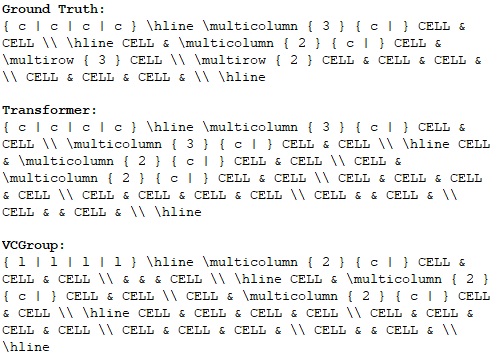} }}%
    \caption{Example failure case of VCGroup method and Transformer-Baseline in the TSR task.}
    \label{fig:tsr_failure_results}%
\end{figure}

\begin{figure}[h]
\centering
    \subfloat[\centering Example 1: Table Image \label{fig:tcr_failure_results(a)}]{{\includegraphics[scale=0.25]{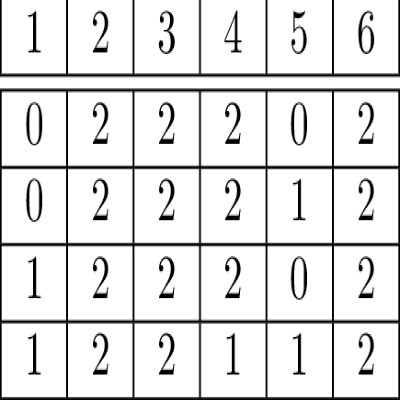} }}%
    \qquad
    \subfloat[\centering Example 1: Outputs \label{fig:tcr_failure_results(b)}]{{\includegraphics[scale=0.5]{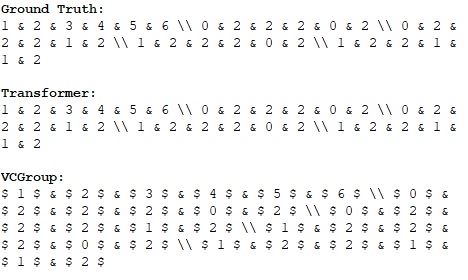} }}%
    \qquad
     \subfloat[\centering Example 2: Table Image \label{fig:tcr_failure_results(c)}]{{\includegraphics[scale=0.25]{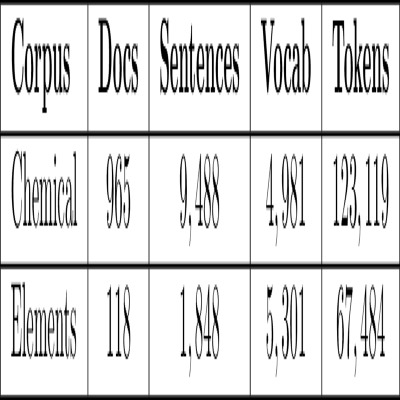} }}%
    \qquad
    \subfloat[\centering Example 2: Outputs \label{fig:tcr_failure_results(d)}]{{\includegraphics[scale=0.5]{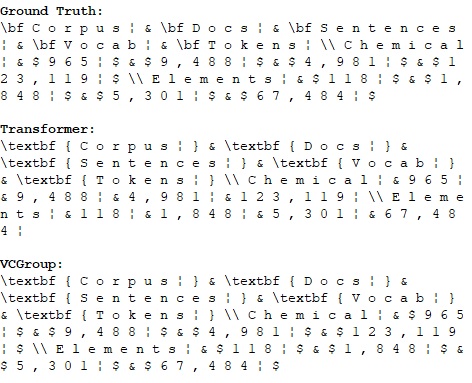} }}%
    
    \caption{Examples of failure cases of VCGroup method and Transformer-Baseline in the TCR task.}
    \label{fig:tcr_failure_results}%
\end{figure}

\textbf{Task I:Table Structure Reconstruction.} In the TSR task, they use the following strategies to improve performance
\begin{itemize}
    \item \textbf{Ranger Optimizer.} Ranger integrates Rectified Adam~\cite{radam}, LookAhead~\cite{lookahead}, and GC (gradient centralization)~\cite{gc} into one optimizer. LookAhead can be considered as an extension of Stochastic Weight Averaging (SWA) in the training stage.

    \item \textbf{Data Augmentation.} Data augmentation methods include shear, affine transformation, perspective transformation, contrast, brightness, and saturation augment is used in our tasks.
    
    \item \textbf{Synchronized Batch Normalization.} Synchronized BatchNorm (SyncBN)~\cite{syncbatchnorm} is an effective type of batch normalization used for multi-GPU training. Standard batch normalization only normalizes the data within each device (GPU). SyncBN normalizes the input within the whole mini-batch. 
    
    \item \textbf{Feature Concatenation of Layers in Transformer Decoder.} Different from the original MASTER model, they concatenate the outputs of the last two transformer layers and apply a linear projection on the concatenated feature to obtain the last-layer feature. The last-layer feature is used for the final classification.
    
    \item \textbf{Model Ensemble.} Model ensemble (by voting and bagging) is a very widely used technique to improve the performance on many deep learning tasks and is used by the participant to improve performance.
\end{itemize}

\textbf{Task II: Table Content Reconstruction.} For the TCR task, ranger optimizer, SyncBN, data augmentation, feature concatenation, and model ensemble are used along with the following additional strategies.

\begin{itemize}
    \item \textbf{Multiple Resolutions.} They experimented with various input image sizes apart from the 400x400 size images provided, e.g., 400×400, 440×400, 480×400, 512×400, 544×400, 600×400.
    \item \textbf{Pre-train Model.} They use a large-scale data set~\cite{table2latex450k} sharing similar objective (table recognition to Latex) with this competition to pre-train their model. To match the target vocabulary, they trim down this dataset to 58,000 samples from the original 450,000 samples.
\end{itemize}

More details about their method can be found in~\cite{he2021pinganvcgroups}.

\section{Results and discussion}
\label{sec:results}
Table~\ref{result_I} and \ref{result_II} shows the results of the TSR and TCR tasks, respectively. VCGroup achieved the highest EM score of 74\% and 55\% for the TSR and TCR task, respectively. 
%Additionally, it performs 2\% better in EM metric vs. the EM @ 95\% metric compared to the Transformer-Baseline for both tasks. 
This shows the method's superior capability in producing better \LaTeX~code sequences. The difference in RA and CA metrics for all the methods is less compared to EM metric, which shows that while VCGroup has superior overall detection, the other methods have similar row and column detection capabilities. VCGroup showcases significant improvement in recognition capability of Alpha-Numeric characters, \LaTeX~tokens, and  non-\LaTeX~ symbols, thus justifying their higher EM score for the TCR task. Figure~\ref{fig:avg_em} presents the average EM scores of the top two performing methods for different lengths of sequences. The plot shows that VCGroup performs significantly better for a longer length of sequences in both tasks. Figure~\ref{fig:tsr_correct_results} and \ref{fig:tcr_correct_results} presents some examples of input tabular image and the corresponding ground truth code, Transformer-Baseline and VCGroup outputs. The examples are chosen to portray the cases where VCGroup outperforms the Transformer-Baseline. For the TSR task, in both examples, the Transformer-Baseline fails to output the last line of the table. For the TCR task, the Transformer-Baseline model fails to predict the dollar (\$) token multiple times and, in the other example, outputs the alternative token \verb+\bf+ while the ground truth contains \verb+\textbf+ to bold text. Figure~\ref{fig:tsr_failure_results} shows an example of a failure case of both VCGroup and Transformer-Baseline in the TSR task. VCGroup incorrectly predicts tabular alignment tokens and the number of columns in the spanning cells. Additionally, it incorrectly predicts \verb+\multicolumn+ token in place of \verb+\multirow+ token. Similarly, the Transformer-Baseline also incorrectly predicts the number of columns in spanning cells and  \verb+\multicolumn+ token in place of \verb+\multirow+ token. Figure~\ref{fig:tcr_failure_results}a and \ref{fig:tcr_failure_results}b shows an example where the VCGroup predicts extra dollar (\$) tokens around the numeric tokens as compared to the Transformer-Baseline. Figure~\ref{fig:tcr_failure_results(c)} and \ref{fig:tcr_failure_results(d)} shows an example where both VCGroup and Transformer-Baseline outputs the alternative token \verb+\textbf+. Additionally, the Transformer-Baseline is not able to correctly predict dollar tokens (\$) as present in the ground truth.

\section{Conclusion} 
\label{conclusion}
This paper discusses the dataset, tasks, and participants' methods for the ICDAR 2021 Competition on Scientific Table Image Recognition to \LaTeX. The participants were asked to convert a given tabular image to its corresponding \LaTeX~source code for this competition.
The competition consisted of two tasks. In task I, we ask the participants to reconstruct the \LaTeX~structure code from an image. In task II, we ask the participants to reconstruct the \LaTeX~content code from an image. The competition winner was chosen based on the highest EM score separately for both the tasks.
The competition portal received a total of 101 registrations, and finally, three teams participated in the test phase of the TSR task, and two teams participated in the test phase of the TCR task. We attribute this low participation to the difficulty of the proposed tasks. In future, we would increase the size of the dataset and include tables from more sources. We would also run the competition for a longer duration to provide the participants ample time to come up with competing solutions.
% Submission by team VCGroup got the highest Exact Match accuracy score of 74\% for Subtask 1 and 55\% for Subtask 2, beating previous baselines by 5\% and 12\% respectively and thus winning the competition.

.

% Although improvements can still be made to recognition capabilities of models, this competition contributes to the development of fully automated table recognition systems by challenging practitioners to solve problems under specific constraints and sharing their approaches; the platform will remain available for post-challenge submissions at \href{https://competitions.codalab.org/competitions/26979}{https://competitions.codalab.org/competitions/26979}.

% The model MASTER~~\cite{DBLP:journals/corr/abs-1910-02562} proposed by the VCGroup team performed better than both CNN baseline as well as transformer baseline. In general, we observe that VCGroup beat the other baselines by significant margin in the TCR task. This clearly indicates the MASTER has more efficacy to learn robust representation towards \LaTeX~ typesetting.\\

% In the future iteration, we plan to add table images with multiple aspect ratios and analyse how these aspect ratios affect the model performance. In this competition we reduce the hardness of the dataset by limiting the vocabulary size with the help of \verb+\LATEX_TOKEN+. In the future version of this competition we aim to increase the hardness of the dataset. 

% \todo{Pratik bhaiya can you add this(why few participants) - argue that competition is hard, few participants got zero accuracy so they tend to drop from the competition.}
% For both the tasks, MASTER~\cite{DBLP:journals/corr/abs-1910-02562} method of VCGroup team produced best results, winning both the tasks. The results of the competition lead us to conclude that scientific table recognition remains a challenging task for state-of-the-art systems.

\section{Acknowledgments}
This work was supported by The Science and Engineering Research Board (SERB), under sanction number ECR/2018/000087.

\bibliographystyle{splncs04}
\bibliography{eacl2021}

\end{document}